\documentclass[runningheads]{llncs}
\usepackage{graphicx}
\usepackage{amsmath,amssymb} 
\usepackage{color}
\usepackage{booktabs}
\usepackage{subfigure}
\begin{document}
\title{CNN-based Real-Time Parameter Tuning for Optimizing Denoising Filter Performance\thanks{Supported by NSERC Discovery Grant and DND Supplement.}}
%
%
\author{Subhayan Mukherjee \and
Navaneeth Kamballur Kottayil \and
Xinyao Sun \and
Irene Cheng}
\authorrunning{S. Mukherjee et al.}
%
\institute{University of Alberta, Edmonton AB T6G2R3, Canada
\email{\{mukherje,kamballu,xinyao1,locheng\}@ualberta.ca}}
\maketitle              
\begin{abstract}
We propose a novel direction to improve the denoising quality of filtering-based denoising algorithms in real time by predicting the best filter parameter value using a Convolutional Neural Network (CNN). We take the use case of BM3D, the state-of-the-art filtering-based denoising algorithm, to demonstrate and validate our approach. We propose and train a simple, shallow CNN to predict in real time, the optimum filter parameter value, given the input noisy image. Each training example consists of a noisy input image (training data) and the filter parameter value that produces the best output (training label). Both qualitative and quantitative results using the widely used PSNR and SSIM metrics on the popular BSD68 dataset show that the CNN-guided BM3D outperforms the original, unguided BM3D across different noise levels. Thus, our proposed method is a CNN-based improvement on the original BM3D which uses a fixed, default parameter value for all images.

\end{abstract}
\section{Introduction}
\label{sec:intro}
Image denoising refers to the process of removing noise from a distorted image to recover the clean image. During acquisition, compression or transmission, images and videos often get corrupted by noise. Thus, when the corruption occurs at a particular stage of the processing pipeline, there is a degradation in quality of output of subsequent steps, ultimately affecting the final visualization. This necessitates the image denoising~\cite{mao2016image} step for signal processing and transmission applications. 

In the real world, accurately predicting the result of noise contamination of a clean signal is difficult, as theoretically, there are innumerable possible noise patterns that can contaminate a clean signal. However, most real-world noise patterns can be approximated by Additive White Gaussian Noise (AWGN), and thus it is commonly discussed in the literature. Consequently, traditional denosing approaches try to model image priors and solve optimization problems, e.g., nonlocal self-similarity (NSS) models~\cite{buades2005non,bm3d2007tip}, sparse representations models~\cite{elad2006image,mairal2009online} and gradient-based models~\cite{rudin1992nonlinear,weiss2007makes}. However, these traditional approaches to denoising are slow due to the optimization process, and thus often unfit for real-time applications. Also, complex and diverse scene content often cannot be denoised effectively using such hand-crafted image priors.

The recent breakthroughs in image denoising come from deep neural networks (DNNs), and especially deep Convolutional Neural Networks (CNNs), which use a discriminative denoising model, e.g., MLP~\cite{burger2012image}, RED-Net~\cite{mao2016image} and DnCNN~\cite{zhang2017beyond}. Their superior performance in many instances is mainly due to the modeling capability of CNNs and the computational capacity of modern GPUs for training progressively deeper and deeper networks. These discriminative models based on deep learning often demonstrate better performance that the traditional model-based methods. However, their performance on unseen data (during inference) often varies depending on the type of data they were trained on. If training data for a particular type of application is not representative enough and the model cannot generalize well enough, the denoising performance will suffer, which is an inherent issue with all learning-based approaches. For natural images, if the test image has been significantly distorted with high noise level, causing most structures and fine details in the original image to get visually obfuscated, the discriminative learning approaches often prove insufficient.

This paper proposes and validates a ``middle ground'' between the above two approaches. It uses the GPU-based implementation of a state-of-the-art model-based approach (namely, the Block Matching 3D filter, BM3D~\cite{bm3d2007tip}) whose parameter is tuned by our proposed CNN in real time, depending on the characteristics of the noisy input image. This approach is ``best-of-both-worlds'' in the sense that its denoising workflow has a well-understood theoretical basis and is thus, fully explainable (BM3D algorithm) unlike end-to-end trained CNNs. At the same time, it optimizes denoising quality by tuning the model parameter using a CNN, which can capture more complex characteristics of the input image than what is possible using traditional hand-crafted methods.

In this paper, we consider a ``non-blind'' denoising scenario like section 5.2.1 of~\cite{blindvsnon}, where the noise is assumed to be AWGN with known standard deviation.

\subsection{Motivation}
As discussed earlier, over the last few decades, the various challenges posed by the denoising problem has been analyzed thoroughly by many researchers and a lot of interesting solutions have been proposed. In the non-learning-based category, one of the greatest and recent breakthroughs was achieved by BM3D. Very recently, researchers have found that BM3D out-performs even deep learning-based methods for real-world, non-AWGN noise, e.g. in photographs captured by consumer cameras~\cite{plotz2017realphotos}. Moreover, efficient GPU implementation of BM3D has significantly improved its time performance~\cite{Honzátko2017}. BM3D has a lot of input parameters which need to be tuned, though most published denoising methods (learning and non-learning based) compare their performance with BM3D using its default parameter values, as mentioned in the original BM3D paper~\cite{bm3d2007tip}. 

In recent years, researchers have experimentally proved that BM3D performance is, in fact, sensitive to its parameter settings and further, that changing some parameter values influence its denoising performance significantly more than changing values of other parameters~\cite{Lebrun2012,Bashar2016}. We repeated those experiments and came to the same conclusion as the researchers that the $\lambda_{3D}$ is one of the few parameters which cause \textit{significant} difference in BM3D's denoising performance.

In BM3D, after grouping of similar (correlated) image blocks (patches), a 3D decorrelating unitary transform is applied to each 3D stack of grouped similar blocks. Enhanced denoising and image detail preservation can only be ensured by choosing a suitable threshold value ($\lambda_{3D}$) for applying a hard thresholding operator on the transform coefficients. This explains why the $\lambda_{3D}$ parameter has significant influence on BM3D denoising quality.

Recently, researchers have tried to adapt the BM3D parameter $\lambda_{3D}$ to the statistical characteristics of the input image and noise~\cite{mrinide2014icassp} using the Noise Invalidation Denoising (NIDe) technique~\cite{nide2010tsp}. However, the parameter $\lambda$ used in NIDe for noise confidence interval estimation has been fixed to the constant value $3$, and the suitability of the method~\cite{mrinide2014icassp} for real-time performance has not been discussed. Researchers have also attempted to adaptively set the distance threshold for grouping similar image blocks, based on the ratio of the mean and standard deviation and the estimated noise intensity~\cite{Dai2013}. Motivated by the observation that the Human Vision System is locally adaptive, in another work~\cite{Egiazarian2017} researchers have tried to vary BM3D parameters according to local perceptual image characteristics in a manner determined by extensive subjective experiments. In yet another work, researchers have tried to incorporate locally-adaptive patch shapes and Principal Component Analysis (PCA) in the 3D transform to improve denoising quality, but at the cost of increasing time complexity many-fold, as well as rendering their algorithm unsuitable for real-time GPU implementation (due to adaptive-shape patches)~\cite{bm3dsapca}. Other researchers~\cite{Bashar2016} have used traditional learning algorithms like Naive Bayes, Support Vector Machine (SVM), K Nearest Neighbors (kNN) and Random Forest to train numerous classifiers to set the $\lambda_{3D}$ value for each block based on the block's texture. However, block-wise prediction of $\lambda_{3D}$ is expected to increase the BM3D time complexity significantly. Yet, the authors did not report the time performance of their proposed method. Also, they used $7\times7$ sized blocks for classification, but did not report or discuss the possible effects of choosing other block sizes. Lastly, even a very recent attempt at replacing parts of the BM3D pipeline with a CNN did \textit{not} show potential for real-time performance, even using the fastest GPUs available in the market~\cite{bm3dnet2018spl}.

In this work, we design a Convolutional Neural Network (CNN) that can predict the $\lambda_{3D}$ parameter value which best denoises a noisy image. We compare the performance of our method by comparing the denoising performance of BM3D (using our CNN-estimated parameter value) against the denoising performance of BM3D using the default value for the $\lambda_{3D}$ parameter, as recommended in the original BM3D paper~\cite{bm3d2007tip}.

\subsection{Our Contribution}
To the best of our knowledge, we are the first to propose a simple, shallow CNN-based real-time solution to predict optimum parameter values for a filtering based denoising algorithm. In this paper, we consider such a state-of-the-art algorithm, BM3D as a use case to demonstrate and validate this proposal. We propose a method that is readily implementable on GPUs and (for our use case) enhances the denoising capability of the recent GPU-based BM3D implementation without significantly increasing the overall time complexity.

The rest of this paper is organized as follows: Related work is given in Section~\ref{sec:related}. We present our proposed method in Section~\ref{sec:proposedmethod}. Experimental results and analysis are in Section~\ref{sec:resdis}. In Section~\ref{sec:confut}, we give the conclusion and future work.

\section{Related Work}
\label{sec:related}

Image denoising is a well studied problem in image processing. Like mentioned earlier, most approaches in the denoising literature rely on modeling image priors ~\cite{bm3d2007tip,elad2006image,rudin1992nonlinear,chatterjee2009clustering}. However, this often leads to over-smoothening of the denoised images (loss of image details) due to incorrect assumptions about the prior.  

The use of non-linear filters is a popular approach in solving image denoising problems. Non-local means based filters are popular examples. Non-local means are a generalization of bilateral filtering which uses photometric distance as a similarity measure~\cite{buades2005non}. BM3D is a further improvement on this scheme where a joint filtering is performed after grouping similar patches from the image. The methods that follow this idea are generally slow but produce good quality results.

Another class of methods to denoise images rely on end-to-end deep learning. These methods rely on training convolutional auto-encoders to convert noisy images to clean images. The neural network learns a set of filters which, when convolved with noisy images, would generate a clean version of the image~\cite{vincent2010stacked,Ye2015StackedAE}. These methods however, often generate images with blurred edges. End-to-end connected networks were limited in their complexity because of the attenuation of gradients in very deep end-to-end frameworks. Neural networks with skip connection were used to solve this issue in Residual network (ResNet)~\cite{he2016deep}. The skip connections help in propagation of gradients and enable deeper layers. This addition led to further improvements to image sharpness after denoising by end-to-end methods. Variations of the residual network formulation have been proposed in ~\cite{mao2016image} and~\cite{zhang2017beyond}. The former uses an encoder-decoder skip-layer connection for faster training and better denoising performance, while the latter adopts the residual learning formulation, but uses identity shortcuts instead of many residual units.

An alternative approach that is used to solve image denoising problems was pioneered by Trainable nonlinear reaction diffusion (TNRD)~\cite{Chen2017TNRD} and Rapid and accurate image super resolution (RAISR)~\cite{romano2017raisr}, which rely on learning a set of structure tensor features to select a filter at each pixel. The filtering and aggregation of the results lead to denoised images with a shallow neural network.

\section{Proposed Method}
\label{sec:proposedmethod}

\begin{figure}[!t]
\centering
\includegraphics[width=0.85\textwidth]{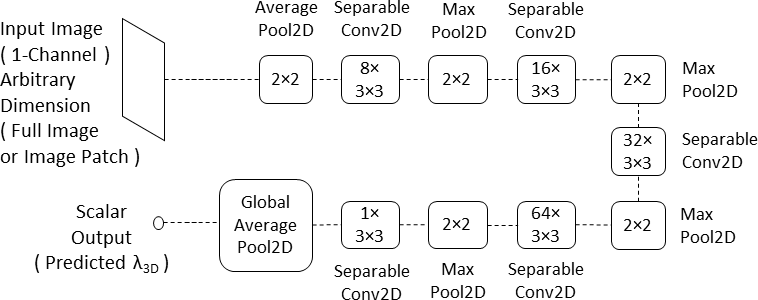}
\caption{Architecture of proposed CNN. Input: Noisy Image; Output: Predicted $\lambda_{3D}$ value. The GlobalAveragePooling2D layer computes the mean of its input and outputs a single scalar value. During training, the cost function is the mean squared error between this value and the optimum $\lambda_{3D}$ value for the input noisy image. The optimum value is determined while preparing the training data.}
\label{fig:model}
\end{figure}

Our literature survey shows that end-to-end learning based denoising performance may be sensitive to training data for particular application domains, whereas most recent improvements to BM3D involve block-based or region-based locally adaptive parameter tuning, which makes them unsuitable for real-time GPU implementation. In our proposed denoising methodology, we train a shallow, fast CNN to predict the optimum value of the BM3D parameter $\lambda_{3D}$ based only on the whole input noisy image. Subsequently, the BM3D algorithm is used to denoise the noisy input image with its $\lambda_{3D}$ parameter set to the value predicted by the CNN. The CNN is trained with pairs of noisy images and the corresponding optimum $\lambda_{3D}$ values. In its current form, our proposed method requires the CNN be re-trained for different AWGN noise $\sigma$ values (we refer to them as noise \textit{levels}). To clarify, the architecture of the CNN remains the same, but the training data and hence, the weights and biases of the trained model are different for different noise levels. However, BM3D itself requires the noise level as an input parameter (``non-blind''), so this is not an extra requirement imposed by our proposed method. As such, the noise level can be estimated following an approach similar to~\cite{Bashar2016}, in which case we can automatically choose the CNN model best suited to that noise level, although this direction has been left as future work. The architecture of the CNN is shown in Fig.~\ref{fig:model}

\setlength{\textfloatsep}{5pt}

In Fig.~\ref{fig:model}, each convolutional layer is represented by a box with rounded corners. Output feature map count is indicated by the integer at the top (1, 8, 16, 32, 64) and filter dimension is indicated at the bottom ($3\times3$). The same representation holds for the non-global pooling layers, except that the number of feature maps remains unchanged in pooling layers, and are thus not explicitly mentioned (pooling window size is mentioned). From the dimensions of the Input layer, one can observe that we do not put constraints on the width or height of the input image or the batch size. In our experiments, we only constrain the input image to be single channel (gray-scale). Thus, we used the CNN to predict $\lambda_{3D}$ values for input images of arbitrary widths and heights. The AveragePooling2D layer right after the Input layer reduces the image dimension and thus the number of convolutions, leading to faster training and inference. We use $3\times3$ separable convolutions~\cite{sifre2014separable} to reduce the number of weights to be trained (for faster convergence). For each of the convolution layers shown in Fig.~\ref{fig:model}, we use the Rectified Linear Unit (ReLU) activation function followed by a MaxPooling2D layer to progressively subsample the feature maps as we move towards the output layer. The pooling window size for all non-global pooling layers in the network is $(2, 2)$. In the GlobalAveragePooling2D (output) layer, we compute the mean of the output of the final convolution layer. The mean is essentially a single scalar value representing the predicted $\lambda_{3D}$ value for the input noisy image. During training, we minimize the Mean Squared Error (MSE) between this mean and the target optimum $\lambda_{3D}$ value for the input image, so that the network can learn to predict the $\lambda_{3D}$ value based on a noisy input image.

\subsection{Design Motivation}

\begin{figure}
\centering

\subfigure[Forest Path]{\label{fig:forest_path}\includegraphics[width=29mm]{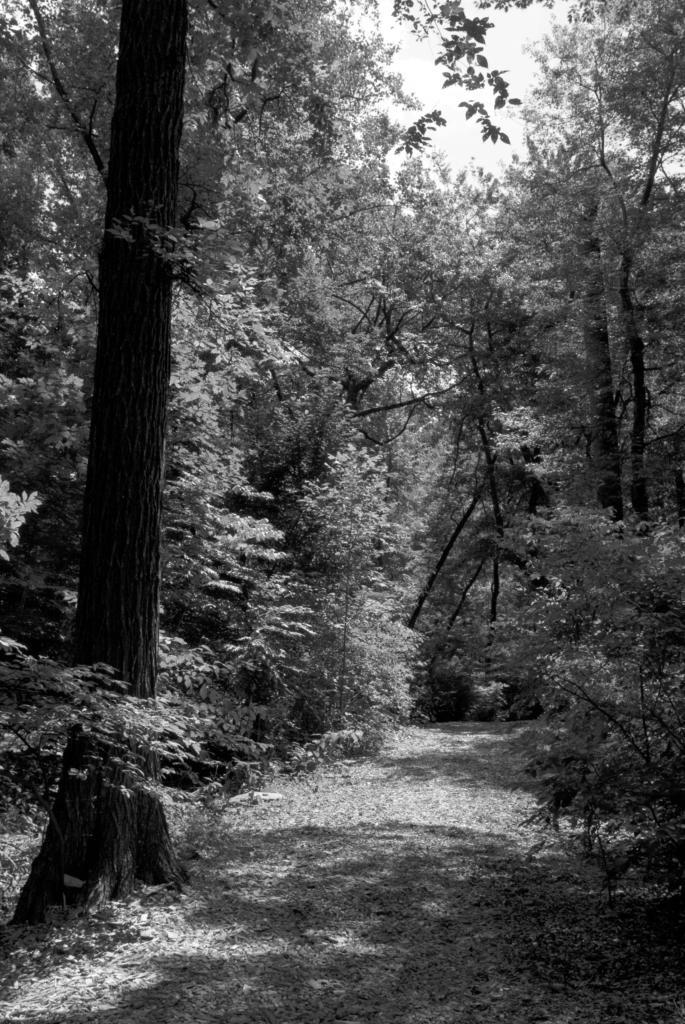}}
\subfigure[Lift Bridge]{\label{fig:lift_bridge}\includegraphics[width=29mm]{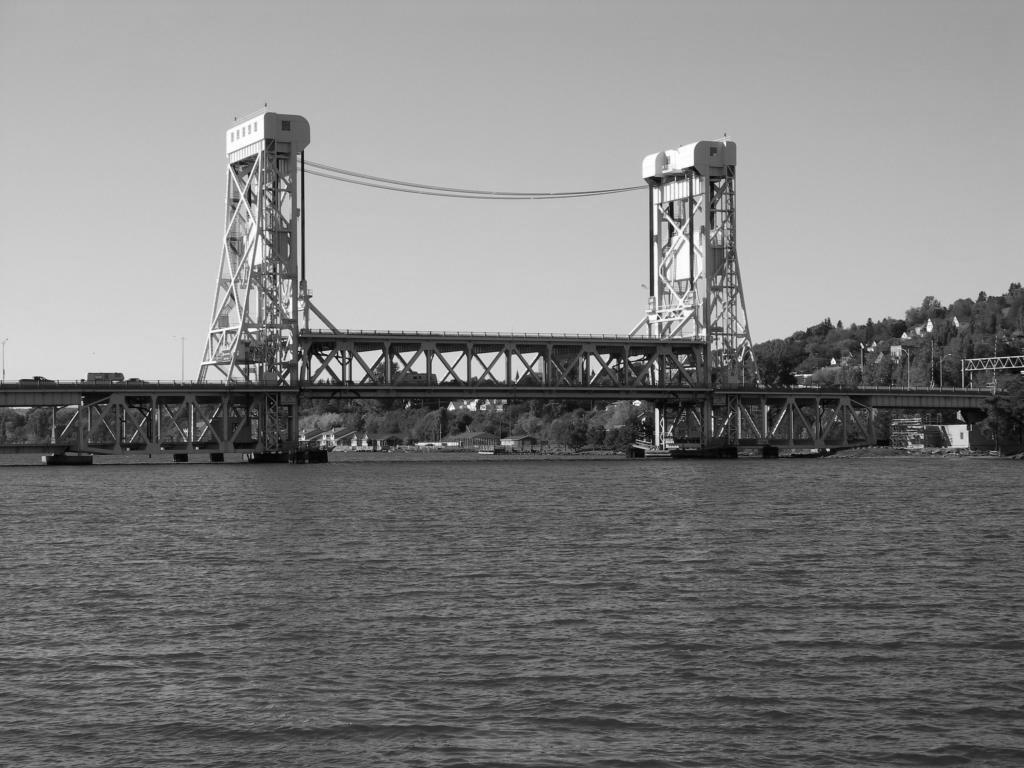}}
\subfigure[Skyscraper]{\label{fig:skyscraper}\includegraphics[width=29mm]{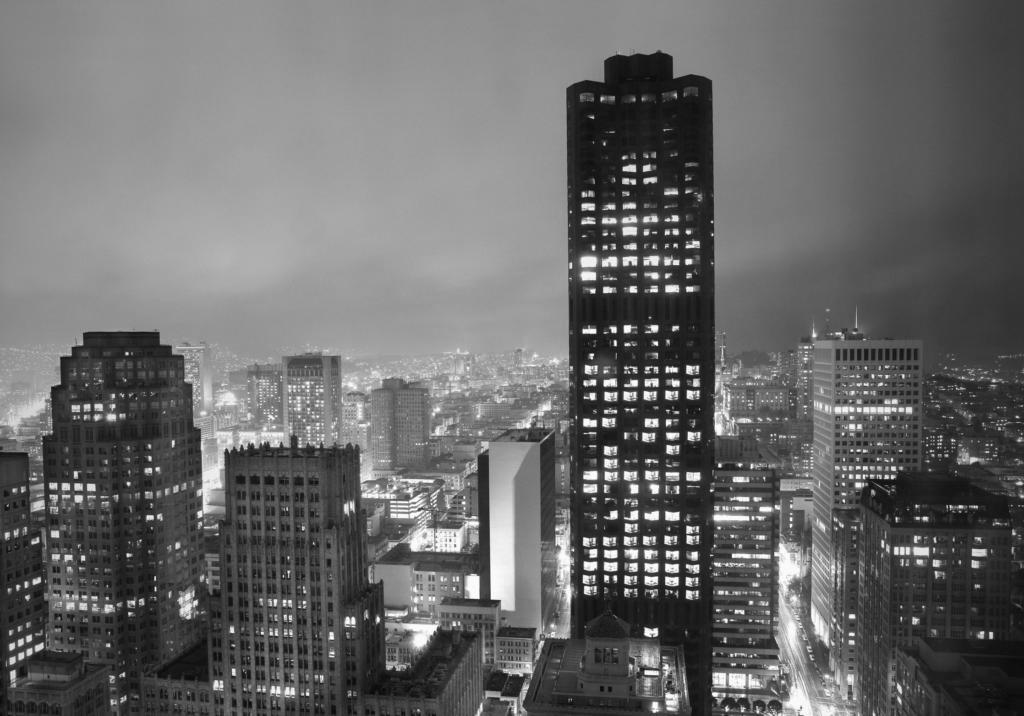}}
\subfigure[Football Stadium]{\label{fig:stadium_football}\includegraphics[width=29mm]{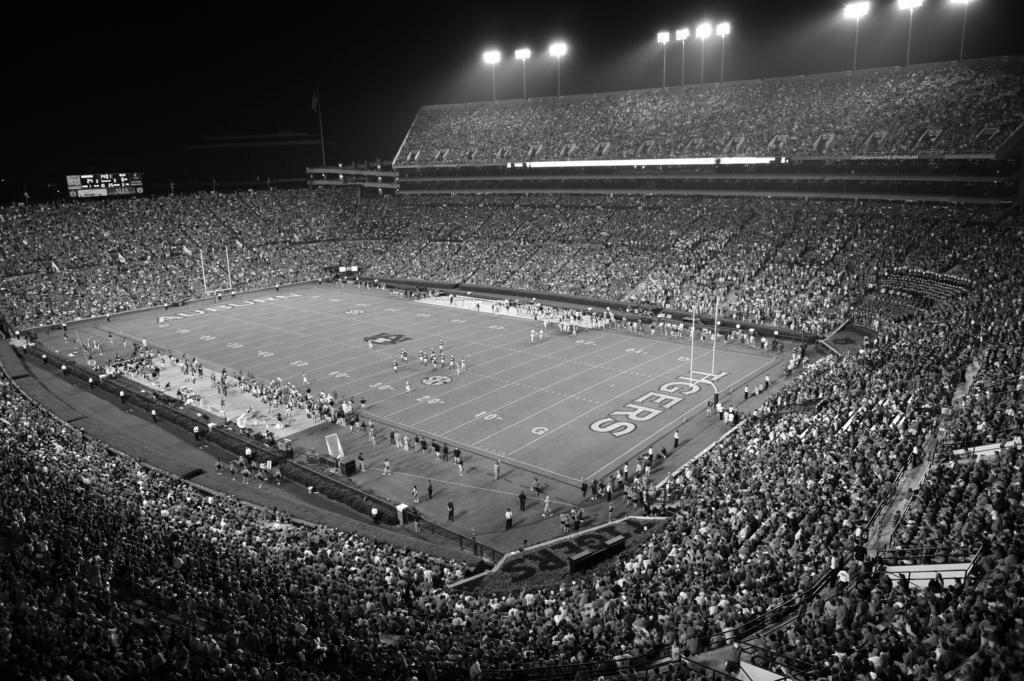}}

\subfigure[MSE Minima ($\sigma=15$)]{\label{fig:forest_path_15}\includegraphics[width=29mm]{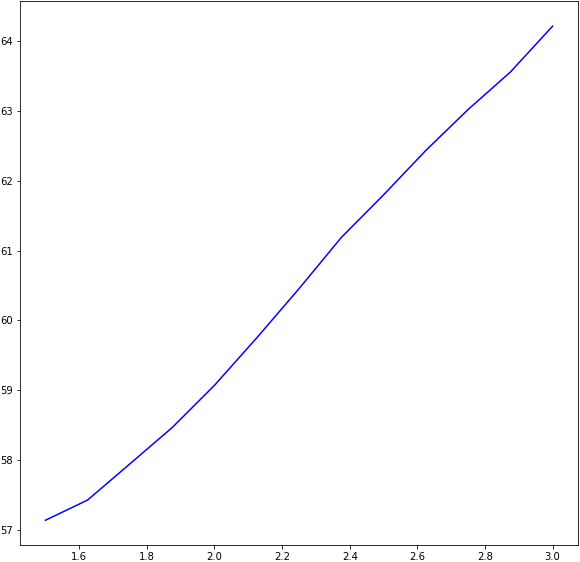}}
\subfigure[MSE Minima ($\sigma=15$)]{\label{fig:lift_bridge_15}\includegraphics[width=29mm]{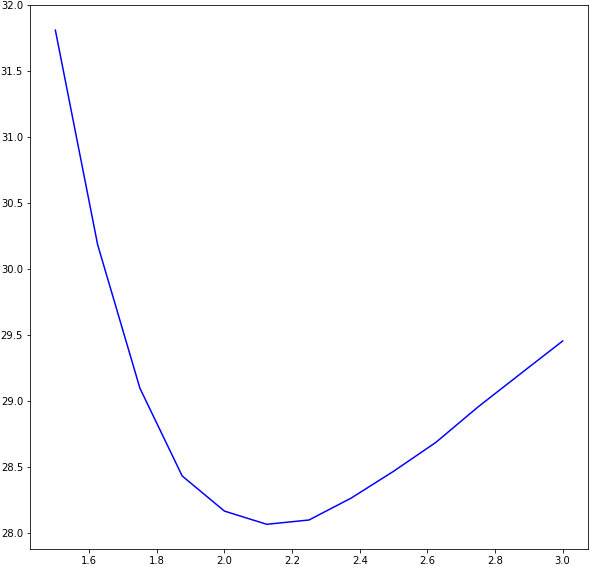}}
\subfigure[MSE Minima ($\sigma=15$)]{\label{fig:skyscraper_15}\includegraphics[width=29mm]{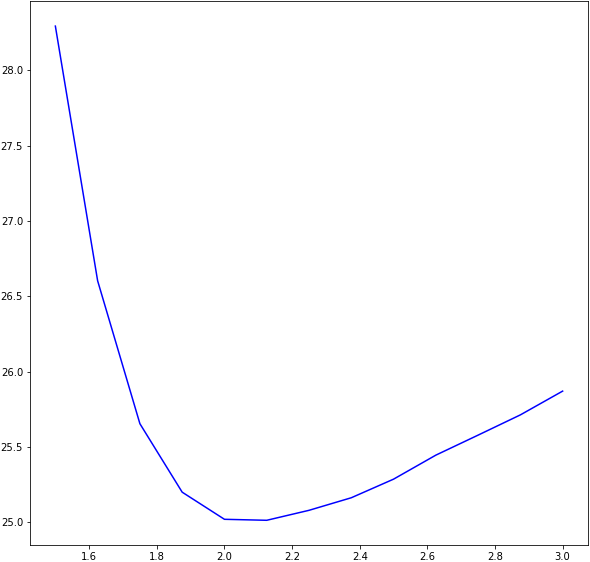}}
\subfigure[MSE Minima ($\sigma=15$)]{\label{fig:stadium_football_15}\includegraphics[width=29mm]{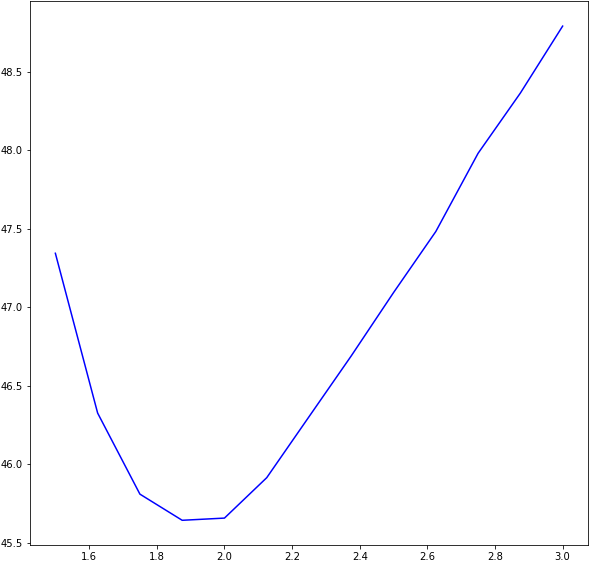}}

\subfigure[MSE Minima ($\sigma=30$)]{\label{fig:forest_path_30}\includegraphics[width=29mm]{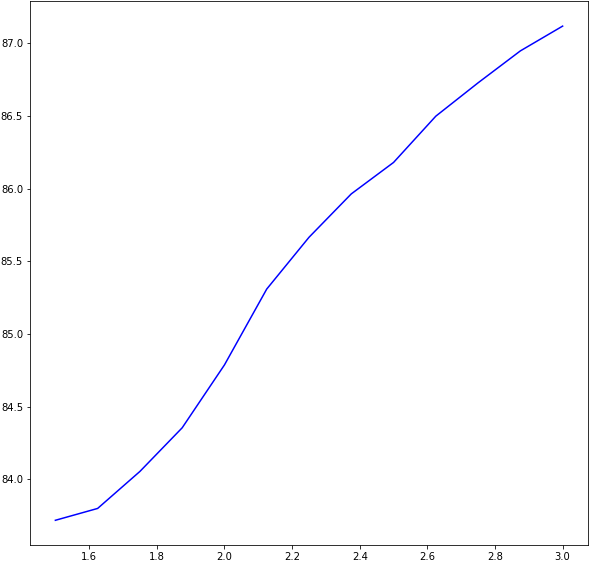}}
\subfigure[MSE Minima ($\sigma=30$)]{\label{fig:lift_bridge_30}\includegraphics[width=29mm]{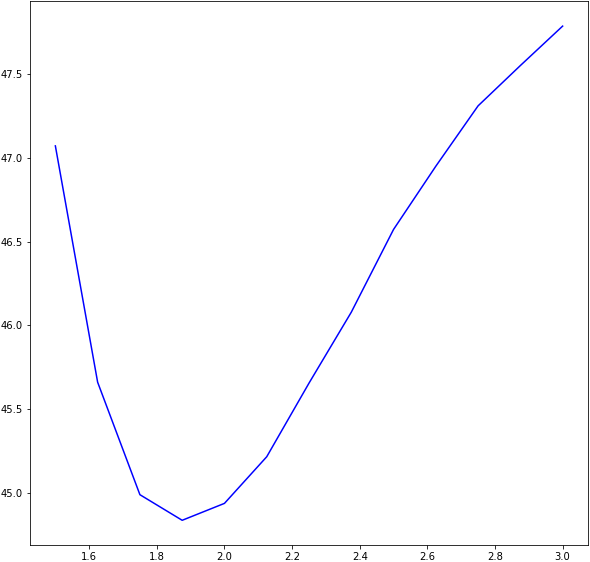}}
\subfigure[MSE Minima ($\sigma=30$)]{\label{fig:skyscraper_30}\includegraphics[width=29mm]{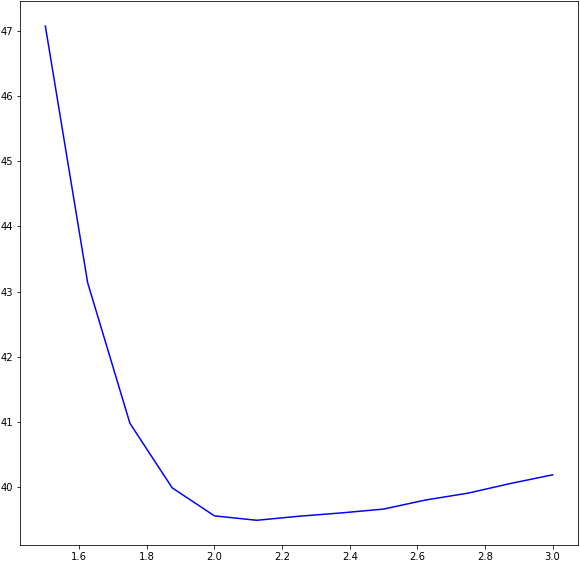}}
\subfigure[MSE Minima ($\sigma=30$)]{\label{fig:stadium_football_30}\includegraphics[width=29mm]{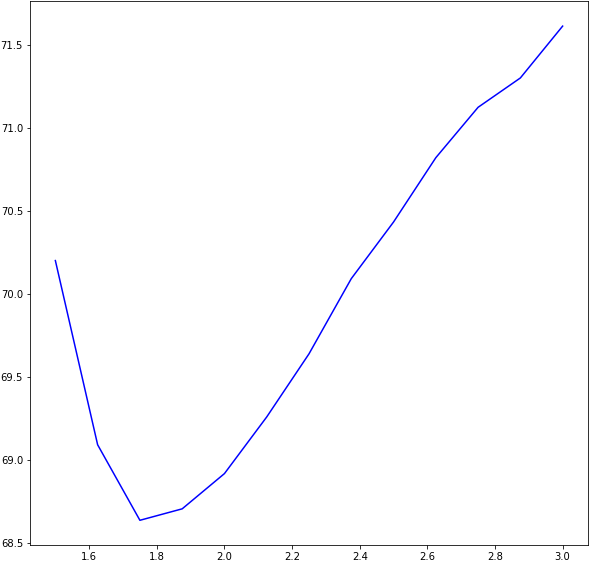}}

\subfigure[MSE Minima ($\sigma=50$)]{\label{fig:forest_path_50}\includegraphics[width=29mm]{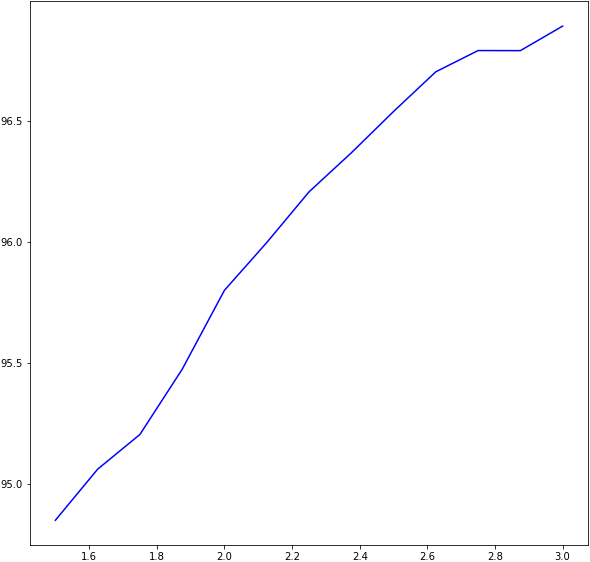}}
\subfigure[MSE Minima ($\sigma=50$)]{\label{fig:lift_bridge_50}\includegraphics[width=29mm]{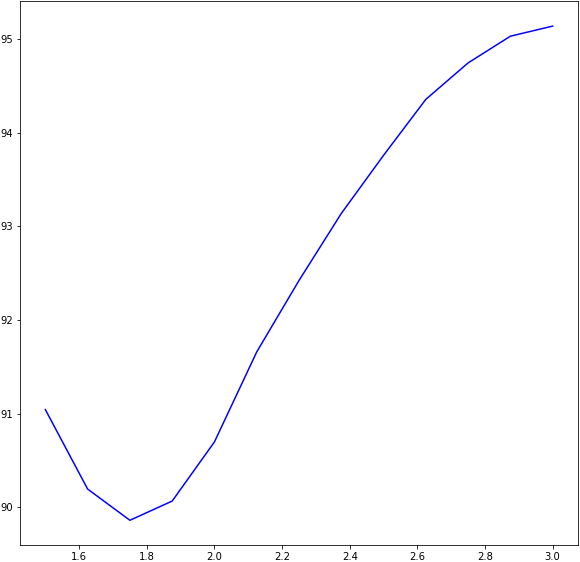}}
\subfigure[MSE Minima ($\sigma=50$)]{\label{fig:skyscraper_50}\includegraphics[width=29mm]{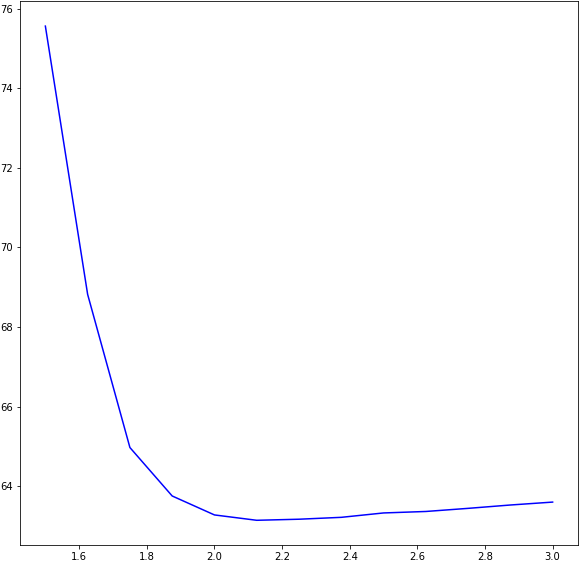}}
\subfigure[MSE Minima ($\sigma=50$)]{\label{fig:stadium_football_50}\includegraphics[width=29mm]{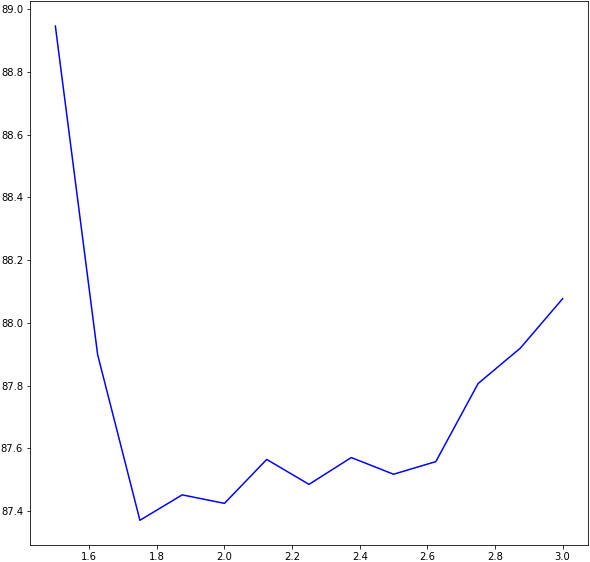}}

\caption{Mean Squared Error (vertical axis) between clean and denoised images of Sun-Hays dataset~\cite{sunhays2012iccp} for different AWGN $\sigma$ values, by varying the $\lambda_{3D}$ parameter (horizontal axis). A distinct minima exists in all cases.}
\label{fig:sunhaysl3d}
\end{figure}

Since the default value for the $\lambda_{3D}$ parameter is $2.7$, we chose different ranges of $\lambda_{3D}$ values for different noise levels, always including the value $2.7$. We observed that when we select the range $(1.0, 3.0)$, we have a minima with respect to MSE between the denoised image and the clean image, across all noise levels. We show few representative results of this experiment on images of the Sun-Hays dataset~\cite{sunhays2012iccp} in Fig.~\ref{fig:sunhaysl3d}. The $\lambda_{3D}$ values are plotted along the horizontal axis and the MSE values along the vertical axis. To increase legibility, the part of the plot corresponding to $\lambda_{3D}$ values less than $1.5$ has been truncated. In the truncated parts, the MSE value was found to display an increasing trend. From all the plots, it can be seen that there is an easily identifiable MSE minima. However, depending on the input image and/or AWGN noise $\sigma$ value, the position of the minima changes. This motivated us to design and train a CNN that could take the noisy image as input and predict the $\lambda_{3D}$ value in order to produce the minimum MSE.

\section{Results and Discussion}
\label{sec:resdis}

\subsection{Training Data Generation}
For training the CNN shown in Fig.~\ref{fig:model}, we have to generate the training data. For this purpose, we first combine two publicly available datasets to create our training dataset:
\begin{enumerate}
\item McGill Calibrated Colour Image Dataset~\cite{Olmos2004} which contains different categories of natural scenes and thus covers a wide range of textures.
\item The ``2017 Unlabeled images [123K/19GB]'' category of the Microsoft Common Objects in Context (COCO) dataset \footnote{http://cocodataset.org/\#download}.
\end{enumerate}
We next converted all the above images to grey-scale (single channel, intensity range: 0-255). We then added AWGN noise with three different levels (15=`low', 30=`medium', 50=`high') to each image. For each noisy version of each image, we reconstructed the noisy image using the BM3D GPU implementation~\cite{Honzátko2017}, by varying the value of the $\lambda_{3D}$ parameter as follows: $[1.0, 1.125, 1.250, ... , 2.875, 3.0]$. Thus, for each image corrupted by each of the three noise levels, we have $17$ reconstructions using different values of the BM3D $\lambda_{3D}$ parameter.
Using each of the three noise levels (15, 30, 50), we compute the MSE of the individual reconstructions with its corresponding clean image, and determine which reconstruction has the minimum MSE (similar to the plots shown in Fig.~\ref{fig:sunhaysl3d}). At each noise level, the $\lambda_{3D}$ parameter value which created the reconstruction having the minimum MSE is assigned as the training label for that noisy image. Thus, the training dataset consists of: 1. \textit{Data}: three versions of an image, each of which is corrupted by a different noise level (low, medium, high), and 2. \textit{Label}: for each version of noisy image, the corresponding $\lambda_{3D}$ parameter value that best denoises it. Thus, from each clean image, we generate three \{noisy image, $\lambda_{3D}$ value\} pairs for inclusion in our training dataset.

\subsection{CNN Training Implementation}
We implement the CNN shown in Fig.~\ref{fig:model} using Keras~\cite{chollet2015keras} with TensorFlow-GPU~\cite{tensorflow2015-whitepaper} back-end. Given a noisy input image (training data), the CNN is trained to minimize the MSE between its predicted output and the ``best'' $\lambda_{3D}$ value (training label) for that noisy input image. Thus, we train the CNN three times (separately), once for each noise level. The computer used for training has Ubuntu 16.04 LTS (64-bit) operating system, Intel Core i7-7700K CPU running at 4.20 GHz ($\times8$ cores), 32 GB system RAM and NVIDIA 1070 GPU with 8 GB GPU RAM. The BM3D GPU implementation took approximately 40-50 milliseconds to generate each training image (for a particular noise level and a particular $\lambda_{3D}$ parameter value). In total, for each noise level, the training data generation took approximately 3 days, but this also includes the time taken to compute the MSEs and determine the minimum MSE for each image. The weights and biases of all CNN layers are initialized by Xavier method~\cite{glorot2010understanding}. The optimization algorithm used is Adam~\cite{kingma2014adam} with $\beta_{1}=0.9$, $\beta_{2}=0.999$, $\epsilon=10^{-8}$. Learning rate was set to $10^{-3}$ with no decay. The network was trained for 50 epochs for each noise level and took approximately $8$ hours for each training session. The images were of varying size ($640\times480$ on average) and orientation (portrait/landscape), but the CNN can take images of any size and orientation, so this was not an issue for us. The batch size was set to $1$ (one image per batch).

\subsection{CNN Performance Evaluation}
\subsubsection{Denoising Parameter Prediction and Quantitative Analysis}
Following the most recent denoising literature, we tested the three trained CNN models (one for each noise level) separately on the Berkeley Segmentation Dataset, containing 68 images (BSD68)~\cite{MartinFTM01}. As mentioned, for each clean image in the dataset, we converted it to grey-scale and added low, medium and high level AWGN. Then, we used the corresponding trained CNN model to predict the $\lambda_{3D}$ parameter value that would best denoise the given noisy input image. We calculated the PSNR and SSIM~\cite{ssimtip2004} for each reconstruction at each noise level for each image. The parameters used to calculate SSIM were the same as those reported in the original paper~\cite{ssimtip2004}, which have been commonly used in the literature. We averaged the PSNR and SSIM for all $68$ images of BSD68 dataset for each of the three noise levels. We also applied BM3D with the default value of the $\lambda_{3D}$ parameter for each test image and each noise level. The results from using our predicted vs. default $\lambda_{3D}$ value in BM3D are summarized in Table~\ref{tab:perfbsd68}. The results show superior performance using our CNN-predicted $\lambda_{3D}$, as compared to default $\lambda_{3D}$ for all metrics across all noise levels.

\begin{table}
\centering
\caption{Performance comparison of proposed CNN-based BM3D parameter prediction}
\label{tab:perfbsd68}
\begin{tabular}{@{}ccccccc@{}}
\toprule
\begin{tabular}[c]{@{}c@{}}Average\end{tabular} & \begin{tabular}[c]{@{}c@{}}MSE\\ (Predicted )\end{tabular} & \begin{tabular}[c]{@{}c@{}}MSE\\ (Default)\end{tabular} & \begin{tabular}[c]{@{}c@{}}PSNR\\ (Predicted)\end{tabular} & \begin{tabular}[c]{@{}c@{}}PSNR\\ (Default)\end{tabular} & \begin{tabular}[c]{@{}c@{}}SSIM\\ (Predicted)\end{tabular} & \begin{tabular}[c]{@{}c@{}}SSIM\\ (Default)\end{tabular} \\ \midrule
\begin{tabular}[c]{@{}c@{}}Noise $\sigma=15$\end{tabular} & 59.06 & 61.42 & 30.93 & 30.78 & 0.8783 & 0.8708 \\
\begin{tabular}[c]{@{}c@{}}Noise $\sigma=30$\end{tabular} & 135.99 & 142.29 & 27.38 & 27.24 & 0.7772 & 0.7669 \\
\begin{tabular}[c]{@{}c@{}}Noise $\sigma=50$\end{tabular} & 255.90 & 267.00 & 24.60 & 24.47 & 0.6650 & 0.6547 \\ \bottomrule
\end{tabular}
\end{table}

\subsubsection{Qualitative Comparison}
\begin{figure}
\centering

\subfigure[Statues (default $\lambda_{3D}$)]{\label{fig:statue_def_15}\includegraphics[width=53mm]{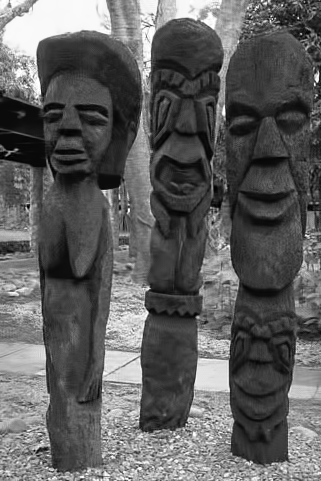}}
\qquad
\subfigure[Statues (prediced $\lambda_{3D}$)]{\label{fig:stat_pred_15}\includegraphics[width=53mm]{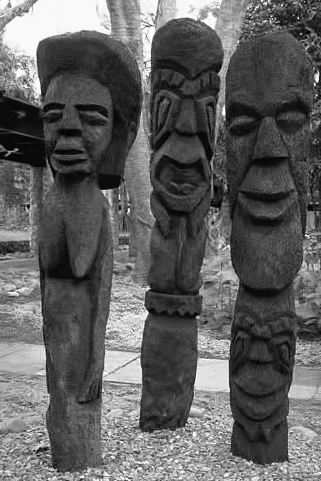}}

\subfigure[Mountains (default $\lambda_{3D}$)]{\label{fig:mountains_def_15}\includegraphics[width=53mm]{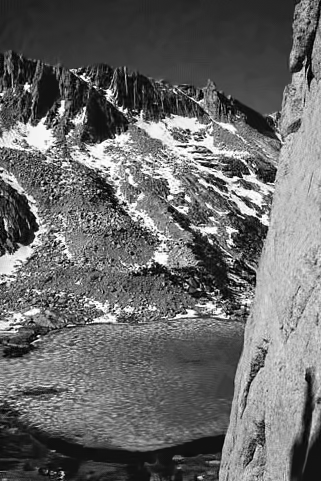}}
\qquad
\subfigure[Mountains (prediced $\lambda_{3D}$)]{\label{fig:mountains_pred_15}\includegraphics[width=53mm]{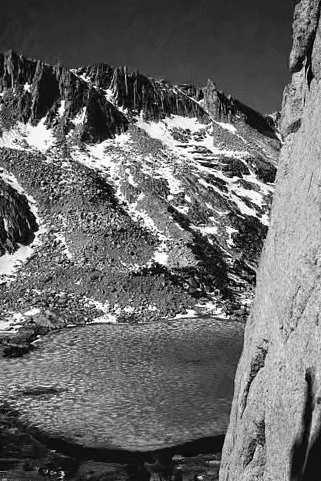}}

\caption{Representative denoised images from the BSD68 dataset. Left: using the fixed, default $\lambda_{3D}$ parameter value ($2.7$); Right: using the CNN-predicted $\lambda_{3D}$ parameter value. Result shown above is for AWGN Noise level $\sigma=15$. Our method better preserves the details in the statues and mountain walls.}
\label{fig:qualcomp_statmont}
\end{figure}

\begin{figure}
\centering

\subfigure[Train (default $\lambda_{3D}$)]{\label{fig:foxface_def_15}\includegraphics[width=40mm]{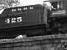}}
\qquad
\subfigure[Train (predicted $\lambda_{3D}$)]{\label{fig:foxface_pred_15}\includegraphics[width=40mm]{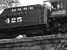}}

\caption{Cropped and zoomed section of a representative denoised image from the BSD68 dataset. Left: using the fixed, default $\lambda_{3D}$ parameter value ($2.7$); Right: using the CNN-predicted $\lambda_{3D}$ parameter value. The denoising result shown above is for AWGN Noise level $\sigma=15$. Our method better preserves the details in the train body's texture.}
\label{fig:qualcomp}
\end{figure}

We compare a cropped part of a representative denoised image using the default $\lambda_{3D}$ parameter value vs. our CNN-predicted $\lambda_{3D}$ parameter value (Fig.~\ref{fig:qualcomp}). We perform similar comparisons for full images in Fig.~\ref{fig:qualcomp_statmont}. From the visual comparison, we can infer that for the most part, the advantage of the parameter prediction lies in greater detail recovery compared to using the fixed, default $\lambda_{3D}$ parameter value ($2.7$). This also explains the higher SSIM score of the proposed method compared to traditional BM3D which uses a fixed $\lambda_{3D}$ parameter value.

\subsection{Analysis and Discussion}
As can be seen from Table~\ref{tab:perfbsd68}, the proposed method of predicting the $\lambda_{3D}$ parameter value based on the input noisy image (instead of using the default value: $2.7$) produces better scores with respect to all three metrics (lower MSE, higher PSNR, higher SSIM). It is important to emphasize that, BM3D scores using default parameter values reported in denoising literature are obtained using the traditional CPU-based implementation of BM3D, which is far much slower than the GPU-based real-time implementation we use~\cite{Honzátko2017}. From Table~\ref{tab:perfbsd68}, the most significant improvement is seen in terms of SSIM score for all noise levels. Since SSIM is a score normalized between $0$ and $1$, even slight increments in SSIM score should be interpreted as noticeable improvements in perceptual image quality.

We found very similar trends as Fig.~\ref{fig:sunhaysl3d} for the BSD68 dataset as well, but we do not reproduce them here for the sake of brevity. Based on our findings, even if the CNN-predicted $\lambda_{3D}$ value is close to the optimal value, the denoising performance using our method is not affected. On the other hand, fixing the parameter value to $2.7$ without taking into consideration the input image characteristics will often produce inferior results, as evident from Fig.~\ref{fig:sunhaysl3d}. It is also worth mentioning that, none of our three trained CNN models ever predicted any $\lambda_{3D}$ parameter value lesser than $1.0$ or greater than $3.0$. That means, the trained networks are quite stable. In fact, even when preparing the training data, we never came across any image for any noise level whose optimum $\lambda_{3D}$ parameter value lies outside the $(1.0, 3.0)$ range.

Lastly, the objective of our work is not to prove the superiority of the proposed method to state-of-the-art learning-based denoising methods which often perform end-to-end learning and are thus subject to the limitations of purely learning-based approaches discussed earlier. Rather, we wish to highlight that even the non-learning based state-of-the-art method BM3D which already has a real-time implementation~\cite{Honzátko2017} can enhance its denoising quality using our proposed real-time, input image based prediction of its $\lambda_{3D}$ parameter value.

\subsubsection{Time Performance}
Our CNN predicts optimum $\lambda_{3D}$ values for all $68$ images of the BSD68 dataset ($481\times321$ resolution) in a total of 0.51 seconds. This translates to an average per-image run time of only 7.5 milliseconds. Our proposed CNN is shallow, uses pooling and separable convolutions. Thus, it runs inference extremely fast, and is ideal for real-time applications when used in conjunction with the real-time BM3D implementation~\cite{Honzátko2017} used in this paper.

\section{Conclusion and Future Work}
\label{sec:confut}
We proposed a novel approach to use CNNs for real-time image-based parameter prediction to enhance the performance of the state-of-the-art denoising algorithm in the non-learning based category, viz. BM3D. Our proposed CNN accepts images of arbitrary size. Experimental results on the popular BSD68 dataset using multiple widely adopted image denoising quality metrics clearly shows that the proposed approach consistently achieves better results than running BM3D with its default, fixed parameter value across different noise levels.

Future work can target predicting other parameters of BM3D or even those of other denoising algorithms using the proposed approach, as well as automatically choosing the most suitable CNN model based on image noise level estimation.
%
%
%
\bibliographystyle{splncs04}
\bibliography{mybibliography}
%




\end{document}